\documentclass[twocolumn,prb]{revtex4}%
\usepackage{graphicx}%
\usepackage{amsmath}%
\usepackage{amsfonts}%
\usepackage{amssymb}

\def\s{{\sigma}}
\def\e{{\epsilon}}
\def\k{{ {\bf k} }}
\def\p{{ {\bf p} }}
\def\q{{ {\bf q} }}
\def\Q{{ {\bf Q} }}

\def\w{{\omega}}
\def\a{{\alpha}}
\def\b{{\beta}}

\def\g{{\gamma}}

\begin{document}
\title{Orbital fluctuation theory in iron-based superconductors:
$s_{++}$-wave superconductivity, structure transition,
and impurity-induced nematic order
}
\author{
Hiroshi \textsc{Kontani}$^{1}$,
Yoshio \textsc{Inoue}$^{1}$,
Tetsuro \textsc{Saito}$^{1}$,
Youichi \textsc{Yamakawa}$^{1}$, and
Seiichiro \textsc{Onari}$^{2}$}
\date{\today }

\address{
$^1$ Department of Physics, Nagoya University and JST, TRIP, 
Furo-cho, Nagoya 464-8602, Japan. 
\\
$^2$ Department of Applied Physics, Nagoya University and JST, TRIP, 
Furo-cho, Nagoya 464-8602, Japan. 
}

\begin{abstract}
The main features in iron-based superconductors would be
(i) the orthorhombic transition
accompanied by remarkable softening of shear modulus, 
(ii) high-$T_{\rm c}$ superconductivity close to the orthorhombic phase, 
and (iii) nematic transition in the tetragonal phase. 
In this paper, we present a unified explanation for them,
based on the orbital fluctuation theory,
considering both the $e$-ph and the Coulomb interaction.
It is found that 
a small $e$-phonon coupling constant ($\lambda\sim0.2$) is enough to 
produce large orbital (=charge quadrupole $O_{xz/yz}$ ) fluctuations,
which causes the $s$-wave superconductivity without sign reversal 
($s_{++}$-wave state).
The derived orbital fluctuations also cause the instability toward
the structure transition due to the 
bound state formation of two orbitons with opposite momenta,
which is called the ``two-orbiton process''.
Moreover, impurity-induced non-local orbital order with $C_2$-symmetry 
is obtained when the orbital fluctuations are strong.
This ``impurity-induced nematic state'' explains the in-plane 
anisotropy of resistivity in detwinned samples.
We stress that (i)-(iii) are reproducible only when
orbital fluctuations with respect to $O_{xz}$ and $O_{yz}$ 
charge quadrupoles are the most divergent.
This fact ensures the reliability of the present model Hamiltonian
and calculation.
\end{abstract}

\sloppy

\maketitle

{keywords: iron-based superconducotrs, orbital fluctuation,
superconductivity, structure transition, shear modulus softening}

\section{Introduction}
\label{sec:intro}

Iron-based high-$T_{\rm c}$ superconductors had been discovered
by Kamiahara {\it et al}. in 2008 \cite{Hosono},
and the highest superconducting (SC) transition temperature $T_{\rm c}$
at present reaches 56K.
The SC state occurs when the crystal structure is tetragonal,
which is realized by chemical doping (or applying the pressure)
on mother compounds, through the structure transition from
orthorhombic to tetragonal.
In the phase diagram, the SC phase is next to the non-SC and 
metallic orthorhombic phase, and the structure transition 
at $T=T_S$ is second-order in Ba(Fe,Co)$_2$As$_2$
\cite{second-order}.
Very large softening of shear modulus $C_S$ suggests 
the existence of strong orbital (quadrupole) fluctuations
\cite{Yoshizawa-old,Yoshizawa,Goto}.
Moreover, the spin-density-wave (SDW) state with 
$\Q\approx(\pi,0)$ occurs in the orthorhombic phase, 
that is, $T_{\rm N}$ is close to but always lower than $T_S$.
These experimental facts suggest a close relation between
the mechanism of superconductivity and 
structure/orbital/SDW transition.

Although the lattice deformation in the orthorhombic phase
is very small ($(a-b)/(a+b)<0.003$),
in-plane resistivity shows sizable anisotropy in the orthorhombic phase.
This fact means that the structure transition is driven by
electron-electron (or electron-optical-phonon) interaction,
not by the cooperative Jahn-Teller effect due to electron-acoustic-phonon
interaction.
Very interestingly, large in-plane anisotropy starts to occur
in the tetragonal phase at $T^*$, 
which is about 10K$\sim$100K higher than $T_S$
in detwinned Ba(Fe,Co)$_2$As$_2$ \cite{detwin}.
The discovery of this ``nematic electronic state'' free from 
lattice deformation had attracted great attention.
The nematic transition also occurs in the tetragonal phase of 
BaFe$_2$(As$_{1-x}$P$_x$)$_2$, confirmed 
by the in-plane anisotropy in the magnetization 
($\chi_a\ne\chi_b$) using the torque measurement under magnetic field
\cite{Matsuda}.
These experimental facts offer us great hints to understand 
the electronic states and the pairing mechanism
in iron-based superconductors.

Now, we have to try to construct a theory
that can explain abovementioned main characters in pnictides
in a unified way, not restricted to the superconductivity.
In the early stage of the study of iron-based superconductors, 
however, many theorists had concentrated on the study of 
pairing mechanism.
Based on spin fluctuation theories,
fully-gapped sign-reversing $s$-wave ($s_\pm$-wave) state had been predicted
 \cite{Mazin,Kuroki,hirschfeld,Chubukov,Tesa}.
The origin of the spin fluctuations is the intra-orbital nesting 
and the Coulomb interaction.
However, the robustness of $T_{\rm c}$ against randomness in iron pnictides 
indicates the absence of sign-reversal in the superconducting (SC) gap
\cite{Onari-impurity,Sato-imp,Nakajima,Li}.
Later, orbital-fluctuation-mediated $s$-wave state 
without sign reversal ($s_{++}$-wave) had been proposed
\cite{Kontani-RPA,Saito,Onari-FLEX}.
The orbital fluctuations mainly originate from the inter-orbital nesting 
and the electron-phonon ($e$-ph) interactions due to 
non-$A_{1g}$ optical phonons.

One of the main merits of the orbital fluctuation scenario is the 
robustness of the $s_{++}$-wave state against impurities or randomness,
consistently many experiments \cite{Sato-imp,Nakajima,Li}.
Moreover, orbital-fluctuation-mediated $s_{++}$-wave state 
scenario is consistent with the large SC gap on the 
$z^2$-orbital band in Ba122 systems \cite{Saito}, 
observed by bulk-sensitive laser ARPES measurement 
\cite{Shimo-Science}.
Note that the ``resonance-like'' hump structure in the neutron 
inelastic scattering \cite{detwin} 
is frequently explained as the spin-resonance due to 
the sign reversal in the SC gap \cite{res-the1,res-the2}.
However, experimental hump structure is well reproduced in terms 
of the $s_{++}$-wave SC state, rather than the $s_\pm$-wave SC state,
by taking the suppression 
in the inelastic scattering $\gamma(\w)$ for $|\w|\le3\Delta$
in the SC state (dissipationless mechanism)
\cite{Onari-resonance,Onari-resonance2}.

In this paper, we present recent developments of 
the orbital fluctuation theory presented in Refs. 
\cite{Kontani-RPA,Saito,Onari-FLEX,Se,Kontani-soft}. 
We present a unified explanation for the 
the following main characters in iron-based superconductors:
(i) orthorhombic transition 
accompanied by remarkable $C_{S}$ softening
\cite{Yoshizawa-old,Yoshizawa,Goto}, and
(ii) emergence of high-$T_{\rm c}$ superconductivity
strong against randomness next to the orthorhombic phase.
We also discuss (iii) ``nematic transition'' in the tetragonal phase
\cite{detwin}
in terms of the impurity-induced non-local orbital order.
It is noteworthy that (i)-(iii) can be explained only when
orbital fluctuations with respect to $O_{xz}$ and $O_{yz}$ 
charge quadrupoles are the most divergent.
This fact assures the reliability of the present theory and model Hamiltonian.

\section{Orbital fluctuations and $s_{++}$-wave superconductivity}
\label{sec:orbital}

\subsection{Antiferro-orbital fluctuations due to inter-orbital nesting}

Here, we study the five-orbital tight-binding model
introduced in Refs. \cite{Kuroki,Miyake}, which reproduces
the experimental multiband structure very well.
We also include both the Coulomb interaction ($U$, $U'$, and $J$)
and the quadrupole-quadrupole interaction 
induced by optical phonons.
In this paper, we introduce the $xyz$-coordinate,
in which $x$- and $y$-axes are along Fe-Fe direction.

Because of the symmetry of As$_4$ tetrahedron,
Fe-ion optical phonon induces the following
quadrupole-quadrupole interaction:
\cite{Saito},
\begin{eqnarray}
H_{\rm quad}=-g(\w_l)\sum_{i}\sum_{\Gamma}^{xz,yz,xy}{\hat O}_{\Gamma}^i {\hat O}_{\Gamma}^i
\label{eqn:Hquad}
\end{eqnarray}
where ${\hat O}_{\Gamma,i}$ is the quadrupole operator 
for channel $\Gamma$ at site $i$ introduced in (I).
$g(\w_l)=g(0)\cdot \w_D^2/(\w_l^2+\w_D^2)$ is the quadrupole coupling
induced by the optical phonons,
were $\w_D=200\sim300$K is the phonon frequency
and $w_l=2\pi T l$ is the boson Matsubara frequency.
Note that ${\hat O}_{\mu\nu}\propto {\hat l}_\mu{\hat l}_\nu
+{\hat l}_\nu{\hat l}_\mu$,
where ${\hat {\bm l}}$ is the angular momentum.
Here, we set $\langle xz|{\hat O}_{yz}|xy\rangle
=\langle yz|{\hat O}_{xy}|xz\rangle
=\langle xy|{\hat O}_{xz}|yz\rangle=1$ by multiplying a constant.
Recently, we have found that Eq. (\ref{eqn:Hquad}) 
is also caused by the multiorbital Coulomb interaction, 
by including the multiorbiton exchange process \cite{mult}
that is absent in the random-phase approximation (RPA)
[S. Onari and H. Kontani, arXiv:1203.28741].

In iron pnictides, antiferro-quadrupole (AFQ)
fluctuations with respect to $\Gamma=xz/yz$
are induced by the quadrupole interaction and the inter-orbital nesting.
The quadrupole susceptibility is 
$\chi_\Gamma^Q(\q,\tau) = \int_0^\beta d\tau e^{i\w_l \tau} 
\langle T_\tau {\hat O}_\Gamma(\q,\tau) {\hat O}_\Gamma(\q,0) \rangle$.

When $g=0$, five quadrupole susceptibilities
$\chi_\Gamma^Q(\q)$ ($\Gamma=xz,yz,xy,z^2,x^2-y^2$) induced by 
Coulomb interaction are small and almost comparable, with peaks
at $\q=(\pi,0)$ or $(0,0)$.
By introducing small quadrupole interaction $g$,
quadrupole susceptibilities with $\Gamma=xz,yz,xy$ are largely enhanced.
The most divergent susceptibilities are
$\chi_{xz}^Q(\q)$ at $\q=(\pi,0)$ and $\chi_{yz}^Q(\q)$ at $\q=(0,\pi)$:
They are approximately expressed as
\begin{eqnarray}
\chi_\Gamma^Q(q)=\frac{c\xi^2}{1+\xi^2(\Q_\Gamma-\q)^2-i\w/\w_0}
 \label{eqn:MMP}
\end{eqnarray}
for $\Gamma=xz$ and $yz$, where 
$\Q_{xz}=(\pi,0)$ and $\Q_{xz}=(0,pi)$.
In two-dimensional systems, orbital fluctuation parameters behave as
\cite{Moriya}
\begin{eqnarray}
\xi^2 &=& l(T-T_{\rm AFQ})^{-1},
 \label{eqn:xi}\\
\w_0 &=& l'(T-T_{\rm AFQ})
\end{eqnarray}
The Weiss temperature $T_{\rm AFQ}$ becomes zero
at orbital quantum-critical-point (QCP). 
In eq. (\ref{eqn:xi}), the unit of the length is 
$a_{\rm Fe-Fe}(\approx 3{\rm Angstrom})$.

In later sections, we will explain that 
the development of $\chi_{xz(yz)}^Q(\Q_{xz(yz)})$
gives rise to not only the $s_{++}$-wave state, 
but also the structure transition as well as 
impurity-induced nematic transition in the tetragonal phase.
Although superconductivity can be caused by other orbital fluctuations
(such as $\Gamma=xy$, $x^2-y^2$, or $z^2$),
both structure and nematic transitions
are uniquely explained only when $\Gamma=xz$ and $yz$ are the most divergent.
Therefore, the present orbital fluctuation model
is considered to be reliable.


\subsection{Phase diagram for $s_{\pm}$- and $s_{++}$-wave states}

\begin{figure}[!htb]
\includegraphics[width=.9\linewidth]{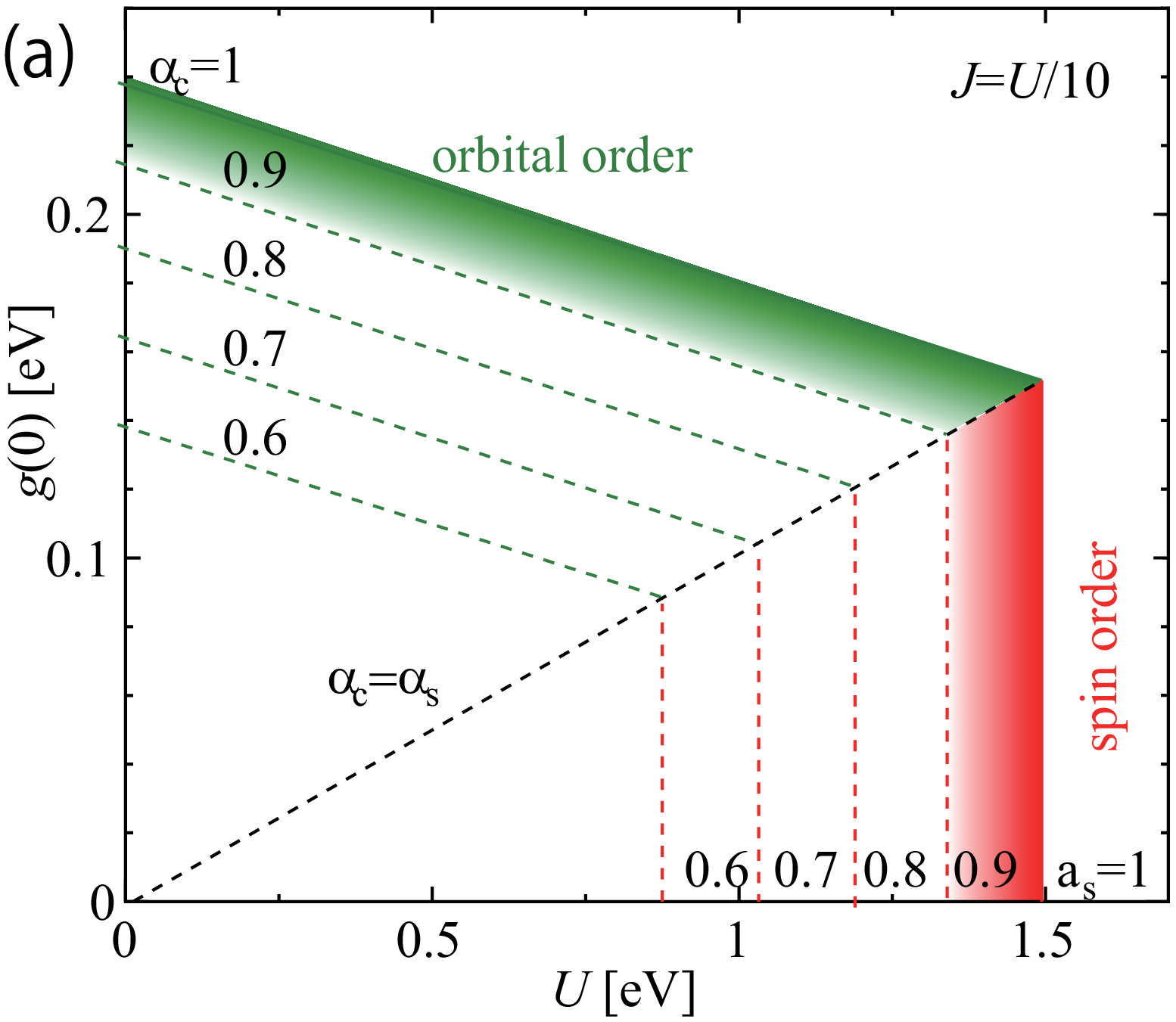}
\includegraphics[width=.99\linewidth]{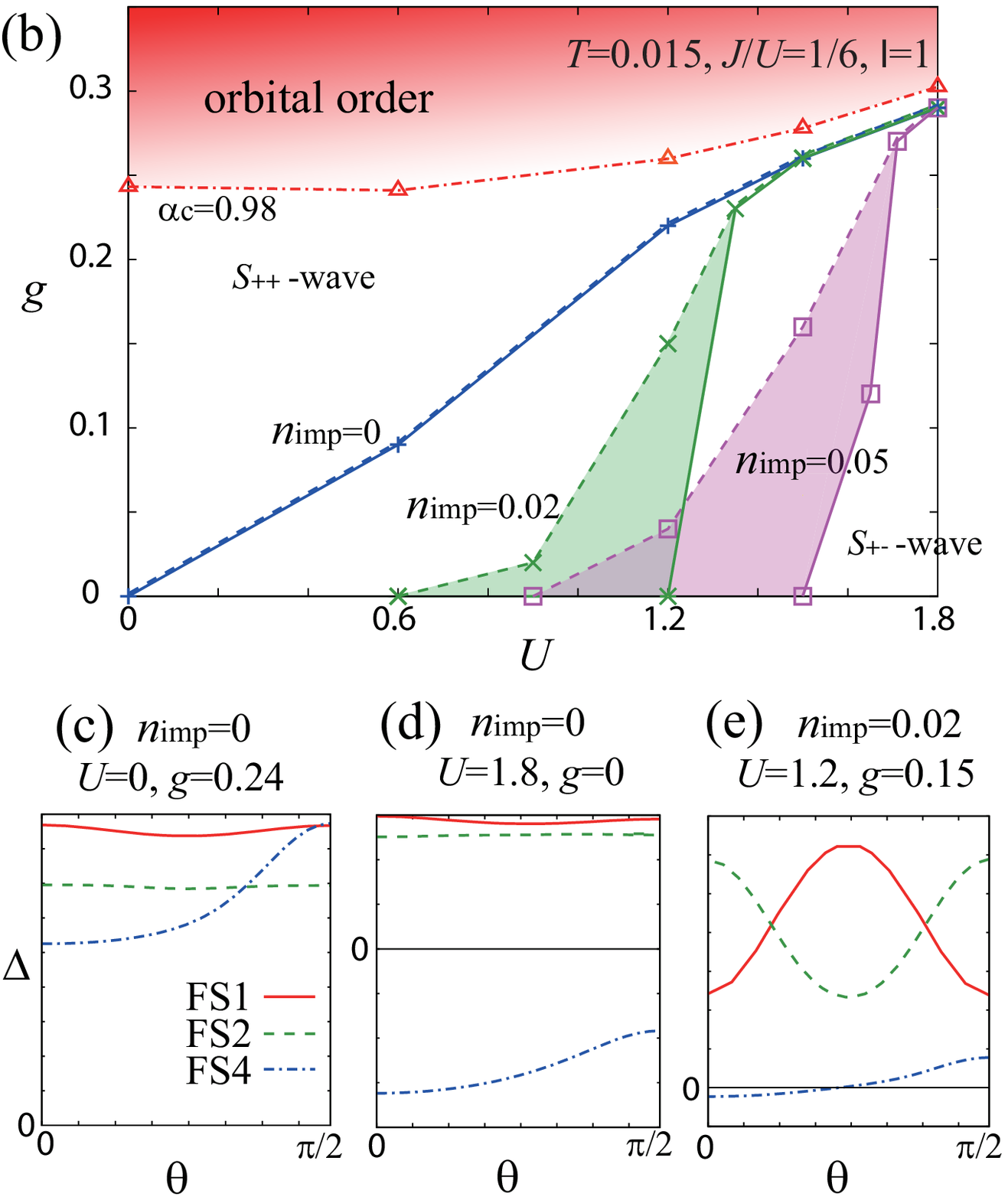}
\caption{
$g$-$U$ phase diagram obtained by (a) RPA for $J/U=1/10$
and (b) FLEX approximation for $J/U=1/6$.
Also, (c)-(e) show the SC gap structure in FS1 (inner h-pocket),
FS2 (outer h-pocket) and FS4 (e-pocket)
for different model parameters.
$\theta_\k={\rm tan}^{-1}((k_y-k_y^0)/(k_x-k_x^0))$,
where $\k^0$ is the center of each FS.
}
\label{fig:phase}
\end{figure}

Figure \ref{fig:phase} (a) shows the phase diagram obtained 
by the RPA (mean-field approximation) for 
the electron filling $n=6.1$,
where $g(0)$ is the quadrupole interaction and 
$U$ is the intra-orbital Coulomb interaction;
we set the Hund's coupling term $J=U/10$ and the
inter-orbital term $U'=U-2J$.
$a_{c}$ ($a_s)$ is the charge (spin) Stoner factor;
$a_{c}=1$ ($a_s=1)$ corresponds to the orbital (spin) ordered state.
For $U=1.3$eV, orbital order appears when $g(0)\approx0.16$eV,
meaning that about 50\% of orbital fluctuations originate from the 
Coulomb interaction.
By solving the SC gap equation, we obtain the $s_{++}$-wave state
when the orbital fluctuations dominate the spin fluctuations 
\cite{Kontani-RPA}.

\

In Fig. \ref{fig:phase} (b), we show the $U$-$g$ phase diagram 
obtained by the FLEX approximation.
The dashed-dotted line represents the condition $\alpha_c=0.98$
at $T=0.015$, corresponding to $g=0.25\sim0.3$.
Therefore, substantial orbital fluctuations emerge
for $\lambda=gN(0)\lesssim0.2$
even if the self-energy correction is taken into account.
On the other hand,
$\alpha_s=0.95$ (0.92) for $U=1.8$ and $g=0$ (0.3)
in the FLEX approximation,
although $U_{\rm cr}=1.25$ for $\alpha_s=1$ in the RPA.
Thus, the renormalization in $\a_s$ is rather larger than that in $\a_c$.
The region of the $s_{++}$-wave state is widely expanded 
in the presence of dilute impurities.
In the present model, both $\chi_{xz(yz)}^Q$ and $\chi_{xy}^Q$ 
are strongly developed.
Since all $t_{2g}$-orbitals on the Fermi surfaces are involved 
in these fluctuations, we obtained a fully-gapped $s_{++}$-wave state
as shown in Fig. \ref{fig:phase} (c),
consistently with many high-$T_{\rm c}$ pnictides.
When the spin fluctuations are comparable to orbital fluctuations,
on the other hand, competition of these fluctuations
induce the nodal structure in the SC gap, as shown in (e).

In summary, we discussed both the
spin-fluctuation mediated $s_{\pm}$-wave state and 
orbital-fluctuation mediated $s_{++}$-wave state in iron-based superconductors.
We have shown that the latter is realized by 
inter-orbital nesting in the presence of small
quadrupole interaction $g$ in eq. (\ref{eqn:Hquad}).
Figure \ref{fig:phase} (b) show the $U$-$g$ phase diagram
obtained by the FLEX approximation.
Since both $s_{\pm}$- and $s_{++}$-states belong to the same A$_{1g}$ symmetry,
a smooth crossover from $s_{\pm}$-state to $s_{++}$-state is realized
as increasing $g$ or impurity concentration $n_{\rm imp}$.
During the crossover, nodal $s$-wave state is realized,
at which the obtained $T_{\rm c}$ is rather suppressed but finite.
(Mathematically, it is impossible to distinguish between 
nodal $s_{++}$- and nodal $s_{\pm}$-states.)
Experimentally, nodal line structure appears in Ba(Fe$_{1-x}$Co$_x$)$_2$As$_2$,
in which both spin and orbital fluctuations are considered to be 
developed \cite{Kasa}.
Recently, we have verified that the 
realization condition for the nodal structure 
is much moderate if we study a realistic ``three dimensional'' model
\cite{Saito-comment}.

\subsection{heavily hole-doped system: KFe$_2$As$_2$}

Here, we study the superconducting state in newly discovered 
heavily h-doped superconductor K$_x$Fe$_2$Se$_2$ ($T_{\rm c}\sim30$K)
based on the ten-orbital model \cite{Se}.
Since the hole-pockets are absent, 
the $s_\pm$-wave state is unlikely.
However, when the Coulomb interaction is large enough,
spin-fluctuation mediated $d$-wave state would appear
due to the nesting between electron-pockets
\cite{Se-d}.
However, the symmetry of the body-centered tetragonal structure
in K$_x$Fe$_2$Se$_2$ requires the existence of nodes in the $d$-wave gap
 \cite{Se},
although fully-gapped $d$-wave state is realized
in the case of simple tetragonal structure.
In the presence of moderate quadrupole interaction $g$,
on the other hand, we find that orbital fluctuations 
give rise to the fully-gapped $s_{++}$-wave state.

\begin{figure}[!htb]
\includegraphics[width=.99\linewidth]{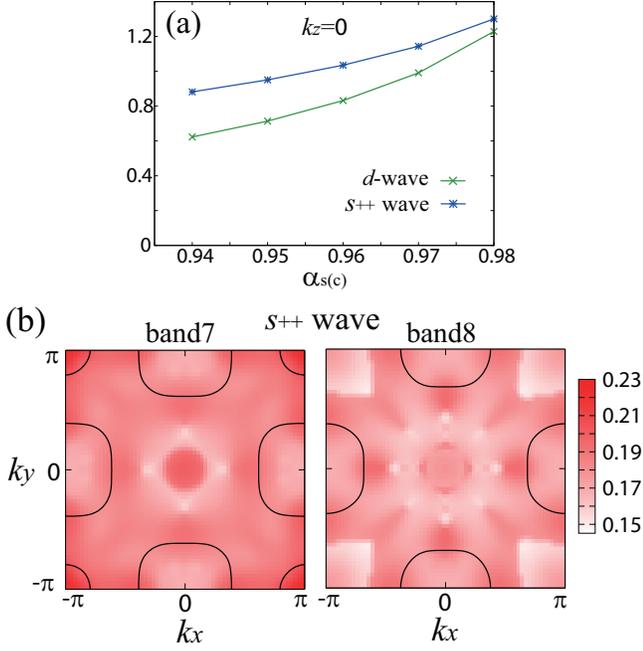}
\caption{
(a) $\a_{s}$- ($\a_c$-) dependence of $\lambda_{\rm E}$
for $d$-wave ($s_{++}$-wave) state at $T=0.03$ eV.
(b) SC gap functions for $s_{++}$-wave state.
}
\label{fig:Se}
\end{figure}

In Fig. \ref{fig:Se} (a), we show the $\a_c$-dependence of 
the eigenvalue of the gap equation, $\lambda_{\rm E}$,
for the $s_{++}$-wave state with $U=0$,
and the $\a_s$-dependence of 
$\lambda_{\rm E}$ for the $d$-wave state with $g=0$.
In calculating the $s_{++}$-wave state,
we use rather larger phonon energy; $\w_D=0.15$ eV,
considering that the calculating temperature is about ten times 
larger than the real $T_{\rm c}$.
The SC gap functions for $s_{++}$-wave state are shown in 
Fig. \ref{fig:Se} (b).
The SC gap would become more anisotropic in case of $U>0$.

Therefore, similar to iron-pnictide superconductors,
orbital-fluctuation-mediated $s_{++}$-wave state
is realized by small $e$-ph interaction $\lambda=gN(0)\sim0.2$.
The obtained $\lambda_{\rm E}$ for the $s_{++}$-wave state
is larger than that for the 
spin-fluctuation-mediated $d$-wave state.
We stress that nodal lines appear on the large e-pockets
in the $d$-wave state, due to the hybridization between two e-pockets
that is inherent in 122 systems,
which is inconsistent with the specific heat measurements 
\cite{HHWen} that report the isotropic SC gap.
We propose that
the study of impurity effect on $T_{\rm c}$ is useful
since nodal $d$-wave state is fragile against impurities.

\section{Orthorhombic structure transition and softening of 
shear modulus $C_S$}
\label{sec:structure}

\subsection{Two-orbiton mechanism}
\label{sec:TO}

In many Fe-based superconductors,
the SC state occurs next to the structure transition,
suggesting a close relation between these two phenomena.
Since the lattice deformation in the orthorhombic phase
is very small ($(a-b)/(a+b)<0.003$),
we can safely rule out the
cooperative Jahn-Teller effect due to electron-acoustic-phonon.
Hereafter, we promise that $a=x$axis and $b=y$ axis.
Below, we discuss the structure transition due to
ferro quadrupole-quadrupole interaction.


For this purpose, we introduce the strain-quadrupole coupling:
\begin{eqnarray}
H_S= \eta_S\sum_i \e_S {\hat \phi}_S,
\label{eqn:HS}
\end{eqnarray}
where $\e_S$ is the strain for the orthorhombic deformation
($\e_S\propto a-b$), ${\hat \phi}_S$ is a quadrupole operator
that belong to the same representation of $\e_S$,
and $\eta_S$ is the coupling constant.
Fernandes {\it et al}. \cite{Fer} considered the 
``spin quadrupole operator'' ${\hat \phi}_S= {\bm s}_1\cdot{\bm s}_2$,
where ${\bm s}_l$ represents the spin operator
at sublattice $l=1,2$.
(In their theory, stripe-type commensurate magnetic correlation is assumed.)
In this case, ${\hat \phi}_S$ is a non-local operator.
On the other hand, the present authors studied the
``charge quadrupole operator'' ${\hat \phi}_S= {\hat O}_{x^2-y^2}$,
which is a local operator.
In the presence of the strain-quadrupole coupling,
the shear modulus $C_S$ is given as
\begin{eqnarray}
C_{S}^{-1}=C_{S,0}^{-1}(1+g_{S}\chi_{S}({\bm 0},0))
\label{eqn:CS}
\end{eqnarray}
where $C_{S,0}$ is the lattice shear modulus, and
$g_{S}=\eta_{S}^2C_{S,0}^{-1}$.
$\chi_{S}(\q,\w)$ is the total quadrupole susceptibility,
given by the Fourier transformation of 
$\chi_{S}(\q,\tau)\equiv
\langle T_\tau {\hat \phi}_\q(\tau){\hat \phi}_{-\q}(0)\rangle$.

Experimentally, $C_S$ follows the Curie-Weiss relation
in the under-doped systems with $T_S>0$:
\begin{eqnarray}
\frac{C_{S}}{C_{S,0}}=\frac{T-T_S}{T-\theta}
\label{eqn:CW}
\end{eqnarray}
On the other hand, $C_{S}$ deviates from the Curie-Weiss law
in the over-doped systems without structure transition.

From now on, we consider the 
charge quadrupole susceptibility for ${\hat \phi}={\hat O}_{x^2-y^2}$,
$\chi_{x^2-y^2}^Q({\bm 0})$, based on the orbital fluctuation theory.
In the RPA, the quadrupole interaction (\ref{eqn:Hquad}) 
together with the Coulomb interaction give rise to the AFQ fluctuations
for the channels $\Gamma=xz,yz$ and $xy$, 
while $\chi_{x^2-y^2}^Q({\bm 0})$ remains small.
Therefore, $C_S$ softening cannot be explained within the RPA.
However, the mode-mode coupling process with respect to $\chi_{xz}^Q(\q)$ 
gives the development of ferro-quadrupole (FQ) fluctuations;
$\chi_{x^2-y^2}^Q({\bm 0})\sim T\sum_\q \{\chi_{xz}^Q(\q)\}^2$.
This process is called the ``two-orbiton term'' in Ref. 
\cite{Kontani-soft}
since the condensation of composite bosons with zero momentum, 
${\hat O}_{xz}(\q){\hat O}_{xz}(-\q)$,
induces the tetragonal structure transition, because of the relation 
${\hat O}_{x^2-y^2} \propto {\hat O}_{xz}^2-{\hat O}_{yz}^2$.
The two-orbiton term has close similarity to
the Aslamazov-Larkin theory of excess conductivity 
given by superconducting fluctuations.



By taking the two-orbiton term,
the irreducible susceptibility is given as
\begin{eqnarray}
{\chi}_{x^2-y^2}^{\rm irr}(\q)=\chi_{x^2-y^2}^{(0)}(\q)+\chi_{\rm TO}(\q)
\label{eqn:bar-TO}
\end{eqnarray} 
where $\chi_{x^2-y^2}^{(0)}(\q)$ is the bare bubble term,
and its temperature dependence is weak.
Within the classical approximation, the two-orbiton term is 
$\chi_{\rm TO}(\q) =T\Lambda^2\sum_\k \chi_{xz}^Q(\q+\k)\chi_{xz}^Q(\k)$,
where $\Lambda$ is the three-point vertex \cite{Kontani-soft}.
The exact expression is given in eq. (61) in Ref. \cite{Kontani-soft}:
\begin{eqnarray}
\chi_{\rm TO}(0) &=& X \xi^2
\left\{ \frac{\w_0}{\pi} \left[ \psi\left(n_{\rm max}+\frac{\w_0}{2\pi T}
+1\right) \right.\right. \nonumber \\
& & -\left.\left. \psi\left(\frac{\w_0}{2\pi T}+1\right) \right] +T\right\}
\label{eqn:chi66-analytic2} ,
\end{eqnarray} 
where $\psi(x)$ is di-Gamma function, and
$\displaystyle X\equiv \frac{(2g)^4 c^2}{4\pi}\Lambda^2$.
($c$ was introduced in eq. (\ref{eqn:MMP}).)
Since $\chi_{\rm TO}(\q)\propto T\xi^2$ for $\q\rightarrow0$,
the two-orbiton term shows strong $T$-dependence.
When $U=0$, the total quadrupole susceptibility is given as
\begin{eqnarray}
{\chi}_{x^2-y^2}^Q(\q)={\chi}_{x^2-y^2}^{\rm irr}(\q)
/(1-g_{S}{\chi}_{x^2-y^2}^{\rm irr}(\q))
\label{eqn:TO}
\end{eqnarray} 
where $g_{S}$ is the quadrupole-quadrupole interaction 
due to acoustic phonon.
As a result, the elastic constant $C_S$ is given by
eqs. (\ref{eqn:CS}), (\ref{eqn:bar-TO}) and (\ref{eqn:TO}),
together with eq. (61) in Ref. \cite{Kontani-soft}.


\begin{figure}[!htb]
\includegraphics[width=.99\linewidth]{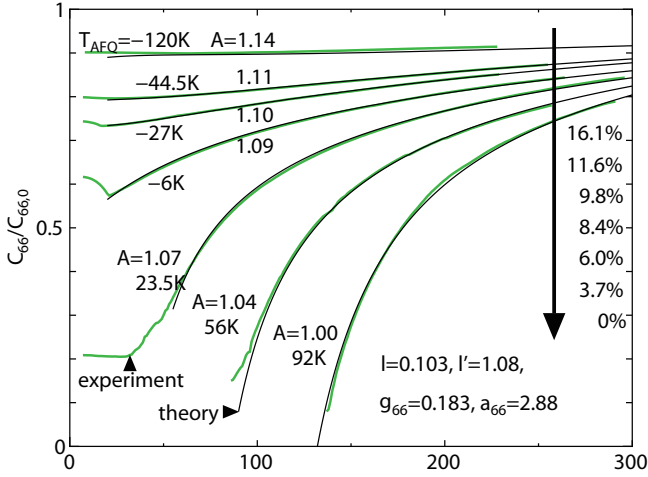}
\caption{
$C_{S}/C_{S,0}$ given by the two-orbiton theory
that reproduce experimental data for Ba(Fe$_{1-x}$Co$_x$)$_2$As$_2$
observed by Yoshizawa \cite{Yoshizawa}.
}
\label{fig:CS}
\end{figure}

In Fig. \ref{fig:CS},
we plot $C_{S}/C_{S,0}$ based on the two-orbiton theory.
We put orbital fluctuation parameters as
$l=1200$K, $l'=1400$K, and $T_{\rm AFQ}=92$K$\sim-120$K.
These values are consistent with theoretical calculation 
based on the FLEX approximation \cite{Onari-FLEX}.
As for the spin fluctuation parameters in cuprates,
$l=0.2$eV (0.1eV) for La$_{1-x}$Sr$_x$CuO$_2$ (YaBa$_2$Cu$_3$O$_7$).
We also set $g_{S}=0.183$eV,
$a_S\equiv \chi_{x^2-y^2}^{(0)}({\bm 0})=2.88{\rm eV}^{-1}$,
and $X\equiv \chi_{\rm TO}(0)/\xi^2T=5.47\sim7.11$.

In the FLEX approximation \cite{Onari-FLEX},
$T_{\rm AFQ}$ changes from positive to negative by carrier doping, while 
other parameters (such as $l$ and $l'$) are insensitive to the doping.
We can fit the recent experimental 
data by Yoshizawa {\it et al.} \cite{Yoshizawa}
for Ba(Fe$_{1-x}$Co$_x$)$_2$As$_2$ with $x=0\sim16$\%, 
by changing $T_{\rm AFQ}$ from $90$K to $-120$K,
together with $X=5.47\rightarrow7.11$.
The fitting data shown in Fig. \ref{fig:CS}
reproduce the experimental data of Ref. \cite{Yoshizawa}
almost perfectly.
This fact is a strong evidence for the success of orbital fluctuation theory
in iron pnictide superconductors.

\subsection{Phase diagram and SC transition temperature}
\label{sec:phase-diagram}

\begin{figure}[!htb]
\includegraphics[width=.99\linewidth]{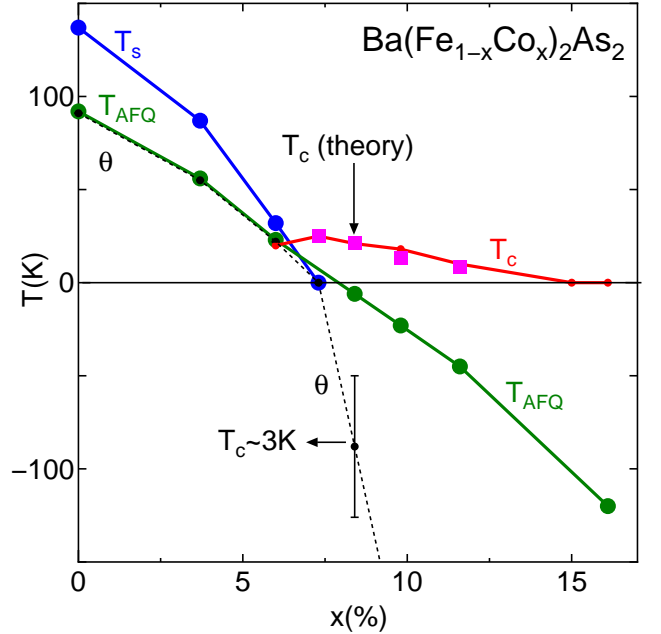}
\caption{
Phase diagram for Ba(Fe$_{x}$Co$_{1-x}$)$_2$As$_2$
obtained by the orbital fluctuation theory.
$T_{\rm AFQ}$ is derived from the temperature dependence of $C_S$
observed in Ref. \cite{Yoshizawa}.
We also show $T_{\rm c}$ derived from eq. (\ref{eqn:gap3}) by solid squares
using orbital fluctuation parameters derived from $C_S$
}
\label{fig:Tc}
\end{figure}

We show the phase diagram for Ba(Fe$_{x}$Co$_{1-x}$)$_2$As$_2$ 
in Fig. \ref{fig:Tc}.
$T_{\rm c}$ and $T_S$ shows the experimental values, and $T_{\rm AFQ}$
is obtained by the fitting using the two-orbiton theory.
While $T_{\rm AFQ}$ should almost coincide to $T_S$ at $T_S=0$ theoretically,
the obtained $T_{\rm AFQ}$ slightly exceeds $T_S$ for $x=7\sim8$\%,
because of the lack of experimental data at low temperatures.
In fact, $\xi^2$ is expected to deviate from the
Curie-Weiss behavior at low temperatures, as predicted by the SCR theory.
In Fig. \ref{fig:Tc}, $\theta$ is obtained experimentally
by using the relation $(1-C_S/C_{S,0})^{-1}\propto T-\theta$
when $C_S/C_{S,0}$ follows the Curie-Weiss relation in eq. (\ref{eqn:CW}).

Based on the obtained orbital fluctuation parameters,
we calculate the SC transition temperature.
Here, we consider the orbital-fluctuation mediated $s_{++}$-wave state 
due to inter-pocket nesting.
The linearized gap equation is given as
\begin{eqnarray}
\lambda_E \Delta_\a(k,\e_n)
&=&\frac{\pi T}{(2\pi)^2}\sum_{\b,p}\sum_{m}\int_\b \frac{d p}{v_{p}^\b}
V_{\a,\b}(k\e_n,p\e_m)
\nonumber \\
& &\times\frac{\Delta_\b(p,\e_m)}{|\e_m|}
\label{eqn:gap1}
\end{eqnarray}
where $\lambda_E$ is the eigenvalue; $\lambda_E=1$ is satisfied at $T=T_{\rm c}$.
$\a,\b$ represent Fermi surfaces (FSs), $k (p)$ is the momentum on 
the FS $\a$ (FS $\b$), and $V_{\a,\b}(k\e_n,p\e_m)$
is the interaction between $(\a,k)$ and $(\b,p)$.
Here, we consider the attractive interaction due to orbital fluctuation;
$V_{\a,\b}\propto \chi_{xz(yz)}^Q$.
For simplicity, we assume that both the Fermi velocity and 
SC gap are isotropic.
Then, the momentum integration along the FS in eq. (\ref{eqn:gap1}) 
can be performed as $\int_\b \frac{d p}{v_{p}^\b}V_{\a,\b}(k\e_n,p\e_m)
=C \xi/(1+|\e_n-\e_m|/\w_0)$, where $C$ is a constant.
If we assume $ \Delta_\a= \Delta_\b$,
we obtain the following simplified gap equation:
\begin{eqnarray}
\lambda_E \Delta(\e_n)
=T\sum_{\b}\sum_{m} \frac{C\xi-\mu^*}{1+|\e_n-\e_m|/\w_0}
\frac{\Delta(\e_m)}{|\e_m|}
\label{eqn:gap3}
\end{eqnarray}
where $\mu^*$ is the Morel-Anderson pseudo potential; we put $\mu^*=0.1$.
We put $C=0.075$ to obtain $T_{\rm c}=25$K when $T_{\rm AFQ}=0$.
In Fig. \ref{fig:Tc}, we show the SC transition temperatures
obtained by eq. (\ref{eqn:gap3}) by solid squares,
which are similar to experimental values.
This fact means that
the orbital fluctuation parameters derived from $C_S$ are reasonable.

\begin{figure}[!htb]
\includegraphics[width=.99\linewidth]{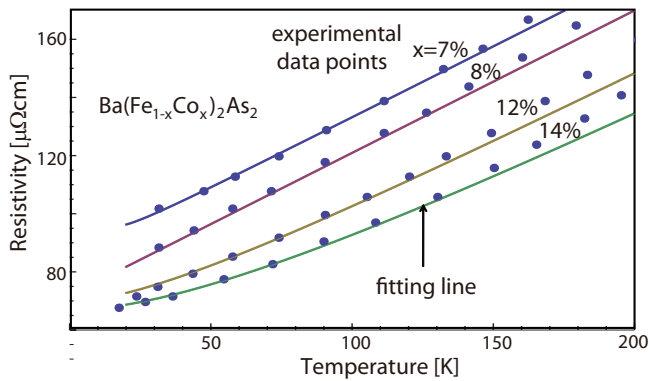}
\caption{
Resistivity given by eq. (\ref{eqn:rho-th}),
using the parameter $T_{\rm AFQ}$ derived from $C_S$.
We can reproduce experimental data 
in Ba(Fe$_{1-x}$Co$_x$)$_2$As$_2$ \cite{Alloul}
below $200$K very well.
}
\label{fig:rho}
\end{figure}

Next, we discuss the non-Fermi-liquid behavior of the resistivity 
caused by the AF orbital fluctuations.
In Fig. \ref{fig:rho}, circles show the resistivity of
single crystal Ba(Fe$_{1-x}$Co$_x$)$_2$As$_2$ reported in Ref. \cite{Alloul}.
According to spin/orbital fluctuation theory \cite{Moriya},
the resistivity in two dimension is given as 
\begin{eqnarray}
\rho=AT^2\xi^2+\rho_0
 \label{eqn:rho-th}
\end{eqnarray}
where $\xi$ is the correlation length of AF fluctuations,
$\rho_0$ is the residual resistivity, and $A$ is a constant.
Using this equation together with $\xi^2=l/(T-T_{\rm AFQ})$ derived from $C_S$,
we can reproduce experimental data below $200$K very well,
as shown in Fig. \ref{fig:rho}.
Here, we put $A=0.49$ for all $x$, and
$(T_{\rm AFQ},\rho_0)=(-8,80)$ for $x=8$\%, 
$(0,72)$ for $x=10$\%, $(-50,70)$ for $x=12$\%,
and $(-90,67)$ for $x=14$\%, respectively.
($T_{\rm AFQ}$'s are obtained from Fig. \ref{fig:Tc}.)
In the present theory,
Fe-ion optical phonons together with the Coulomb interaction
induce the AF-orbital fluctuations, which give rise to 
the $s_{++}$-wave state and non-Fermi-liquid-like transport phenomena.
Moreover, two AF-orbitons with zero total momentum
induce the ferro-orbital fluctuations, which are the origin of the
orthorhombic structure transition.
Therefore, the present orbital fluctuation theory can explain 
both the structure transition and the superconductivity.

\subsection{Difficulties in other orbital fluctuation theories}
\label{sec:ferro}

Here, we discuss various difficulties in 
other orbital fluctuation models proposed for 
iron-based superconductors \cite{Yanagi1,Yanagi2}
To explain the orthorhombic structure transition,
the ferro-$O_{x^2-y^2}$ fluctuations should be the most divergent.
The $U'>U$ model in Ref. \cite{Yanagi1}
cannot explain the structure transition 
since the most divergent fluctuation is $O_{z^2}$-type.

\begin{figure}[!htb]
\includegraphics[width=.99\linewidth]{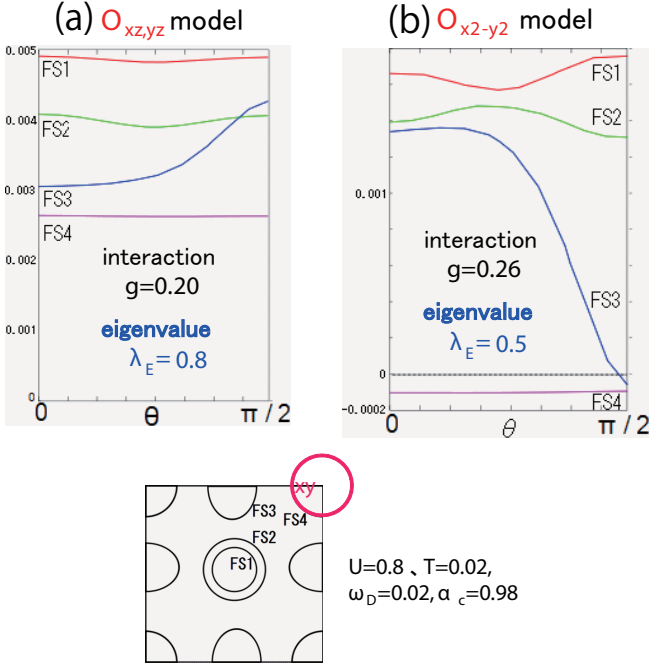}
\caption{
The SC gap functions for $\alpha=0.98$ in the 
(a) $O_{xz/yz}$-fluctuation model and
(b) $O_{x^2-y^2}$-fluctuation model.
In (b), the $xy$-orbital hole-pocket around $(\pi,\pi)$
is almost gapless, since $xy$-orbital is not involved in 
the $O_{x^2-y^2}$-fluctuations.
In (a), the eigenvalue of the gap equation (\ref{eqn:gap1}),
$\lambda_E$, is larger with smaller interaction $g$.
}
\label{fig:gap-F-AF}
\end{figure}

Next, we consider the "$O_{x^2-y^2}$-fluctuation model, 
in which all quadrupoles other than $O_{x^2-y^2}$ do not fluctuate.
Then, electrons with $xy$-orbital character,
which occupies one-third of the total DOS at Fermi level,
are not involved in the $O_{x^2-y^2}$-fluctuations,
since $\langle xy|{\hat O}_{x^2-y^2}|m\rangle=0$
for any $d$-orbital $m$.
This fact means that 
``gapless SC state with large residual DOS'' 
is realized in the $O_{x^2-y^2}$-fluctuation model.
Figure \ref{fig:gap-F-AF} shows the gap structure given by 
the (a) ``$O_{xz/yz}$-fluctuation model'' in Ref. \cite{Kontani-RPA} 
and (b) ``$O_{x^2-y^2}$-fluctuation model'' in Ref. \cite{Yanagi2}. 
In the latter model, the quadrupole interaction
$H'=-g(\w)\sum_i {\hat O}_{x^2-y^2}^i\cdot {\hat O}_{x^2-y^2}^i$
is introduced, considering the As-ion acoustic phonons.
In (a), fully-gapped $s_{++}$-wave state is realized.
In (b), in contrast,
the SC gap on the $xy$-orbital hole-pocket around $(\pi,\pi)$
in the unfolded BZ is almost gapless, and the SC gap on e-pockets
is highly anisotropic.
Therefore, $O_{x^2-y^2}$ fluctuation model cannot explain 
the fully-gapped nor nodal gap structure observed in 
almost all iron-based superconductors.

We also comment on the ``$T$-linear resistivity'' observed in 
various iron-based superconductors.
In the ferro-$O_{x^2-y^2}$ fluctuation model,
the portion of Fermi surfaces with $xy$ orbital
would become the cold-spot with $\gamma\propto T^2$.
Thus, a conventional Fermi liquid behavior $\rho\propto T^{2}$ 
would be obtained in the ferro-$O_{x^2-y^2}$ fluctuation model,
inconsistently with experiments.

Therefore, orbital fluctuations theories in Refs. \cite{Yanagi1,Yanagi2}
have serious difficulty 
in explaining the ``fully-gapped'' $s_{++}$-wave state 
as well as non-Fermi liquid transport phenomena
in iron-based superconductors.

\subsection{Effective Action}
\label{sec:act}

In Ref. \cite{Kontani-soft},
we have calculated the shear modulus $C_S$
using the Green function method.
Since the derivation was rather complex,
we rederive the same expression in a more simple manner
by introducing the Hubbard-Stratonovich field,
in analogy to the analysis in Ref. \cite{Fer}.
By performing the Hubbard-Stratonovich transformation,
we introduce the following effective action
$S=S_0+S_{\rm int}+S_{\rm strain}$
for the charge quadrupole field $\phi_\Gamma$ ($\Gamma=XZ,YZ$):
\begin{eqnarray}
S_0[\phi_\Gamma]&=&\frac12\sum_{\Gamma}^{XZ,YZ}\int_q 
\left\{ \chi_\Gamma^{Q}(q) \right\}^{-1}\phi_\Gamma^2(q),
 \label{eqn:S0}\\
S_{\rm int}[\phi_\Gamma]&=&-\frac{g_{S}'}{2}\int_x \{\phi_{XZ}(x)\phi_{YZ}(x)\}^2
 \label{eqn:Sint}\\
S_{\rm strain}[\phi_\Gamma]&=&\eta_S' \int_x \e_{S}(x)\phi_{XZ}(x)\phi_{YZ}(x)
 \label{eqn:Sst}
\end{eqnarray}
where $\int_x\cdots = \int_0^{1/T}d\tau \int d^2x \cdots$,
$S_0$ represents the most divergent AF quadrupole susceptibility 
$\chi_{XZ(YZ)}^Q(q)$, $S_{\rm int}$ is the quadrupole-quadrupole interaction
due to acoustic phonons, and 
$S_{\rm strain}$ is the strain-quadrupole coupling.
In this subsection,
we introduce the $XY$-coordinate that is $-45$ degree rotated from the 
$xy$-coordinate along $z$-axis.
Apparently, $\phi_{XZ}=(\phi_{xz}-\phi_{yz})/\sqrt{2}$,
$\phi_{YZ}=(\phi_{xz}+\phi_{yz})/\sqrt{2}$, and $O_{x^2-y^2}=O_{XY}$.
In Eqs. (\ref{eqn:Sint}) and (\ref{eqn:Sst}), 
$g_{S}'\equiv g_{S} \cdot \Lambda^2$ and $\eta_S'\equiv\eta_S \cdot \Lambda$.
where $\Lambda$ is the three-point vertex with respect to
$(O_{XY},O_{XZ},O_{YZ})$ that had been analyzed in Ref. \cite{Kontani-soft}.

Now, the quadrupole susceptibility 
$\chi_{x^2-y^2}^Q$ is given by the second derivative of the 
partition function with respect to $\eta\e_S(x)$.
If we put $g_S'=0$, the quadrupole susceptibility is given as
$\Lambda^2 \chi_{XZ}^Q(x)\chi_{YZ}^Q(-x)$.
By performing the Fourier transformation
and taking the interaction $g_S'$ into account, 
the total quadrupole susceptibility is obtained as
\begin{eqnarray}
\chi_{x^2-y^2}^Q(q)&=& \Lambda^2\frac{{\chi^{\rm irr}}'(q)}{1-g_S'{\chi^{\rm irr}}'(q)},
 \label{eqn:chix2y2} \\
{{\chi}^{\rm irr}}'(q)&=& \int_k \chi_{XZ}^Q(k+q)\chi_{YZ}^Q(q)
\nonumber \\
& &+ \Lambda^{-2}{\chi}_{x^2-y^2}^{(0)}(q)
 \label{eqn:barchix2y2}
\end{eqnarray}
where $\int_q\cdots = T\sum_n \int \frac{d^2q}{(2\pi)^2} \cdots$.
Here, we have dropped both the vertex and self-energy corrections.
Apart from the factor $\Lambda^2$,
the first term the irreducible susceptibility
(\ref{eqn:barchix2y2}) is equivalent to 
${\chi}_{\rm TO}$ in eq. (\ref{eqn:chi66-analytic2}),
or $\chi_{0,{\rm nem}}$ in Ref. \cite{Fer}.
We also added the bare susceptibility ${\chi}_{x^2-y^2}^{(0)}$
as the second term in eq. (\ref{eqn:barchix2y2}).
Due to the two-orbiton term,
${\chi^{\rm irr}}'({\bm 0})$ is strongly enhanced near AFQ-QCP
in proportion to $\xi^2$.
When $\g_S'$ is finite, $\chi_{x^2-y^2}^Q({\bm 0})$ diverges
even if $\xi$ is finite,
and therefore the orthorhombic structure transition takes place in 
iron-based superconductors.

Note that the present action in eqs. (\ref{eqn:S0})-(\ref{eqn:Sst})
are mathematically equivalent to eqs. (3) and (4) in the spin-quadrupole 
theory in Ref. \cite{Fer}, 
by replacing $\phi_{\Gamma}$ ($\Gamma=XZ,YZ$) with $\phi_{i}$ ($i=1,2$),
and $\chi_\Gamma^Q(q)$ with $\chi^s(q)$.
Thus, the derived $T$-dependences of $C_S$ are essentially the same.
However, the coupling between strain and spin-quadrupole, $\eta_S$,
would be too small to fit experimental data:
We will discuss this issue below.

\subsection{Comparison with spin quadrupole theory}
\label{sec:comparison}

As discussed in Sec. \ref{sec:structure},
the orthorhombic structure transition is cause by 
the divergence of the FQ susceptibility with ${\hat x}^2-{\hat y}^2$ symmetry.
In this paper, we discussed the FQ fluctuation
induced by the AF quadrupole fluctuations, 
due to the two-orbiton process.
That is, antiferro-orbital fluctuations 
with respect to $O_{XZ}$ and $O_{YZ}$ (=orbitons) induce the $s_{++}$ wave 
superconducting state, and the bound-state formation of two-orbitons with 
zero momentum, $O_{XZ}(\q)O_{YZ}(-\q)\sim O_{XY}({\bm 0})$, give rise to 
the development of $\chi_{XY}^Q(0)=\chi_{x^2-y^2}^Q(0)$.

In this subsection, we discuss the ``spin-nematic theory'' for the 
structure transition discussed in Refs. \cite{Kiv,Fer},
and explain a close relation to the two-orbiton mechanism.
Their theories can be interpreted as
the ``two-magnon process'' in our language.
They had studied the non-local spin quadrupole operator
${\hat \phi}_S={\bm m}_1\cdot{\bm m}_2$ as explained in Sec. \ref{sec:TO},
and shown that the Ising-like order ${\hat \phi}_S\ne0$
In both theories, the Ising-like order 
($\langle {\hat \phi}_S \rangle \ne0$) 
occurs prior to the vector order 
($\langle {\bm m}_i \rangle,\langle (O_{xz},O_{yz}) \rangle \ne0$), 
since the latter is easily suppressed by thermal and quantum fluctuations.
When the boson is orbiton (magnon), the realized superconductivity is the 
$s_{++}$ ($_\pm$) wave state.

According to the fitting done in Sec. \ref{sec:structure},
we have to assume $g_S\sim O(0.1)$eV to reproduce
$E_{\rm JT}\equiv T_S-\theta \sim O(10)$K;
the corresponding dimensionless $e$-ph coupling is $\lambda\lesssim0.1$.
In this case, the required strain-quadrupole interaction $\eta_S$
is about $0.5$eV/Angstrom.
As for the strain-charge-quadrupole interaction,
$\eta_S^{\rm charge}\sim 0.5$eV is actually obtained by the point-charge model
 \cite{Kontani-RPA}.
On the other hand, the strain-spin-quadrupole interaction $\eta_S^{\rm spin}$
would be of order
$\sim \delta J(R)|m|^2/\delta R$, where $J(R)$ is the 
nearest-neighbor magnetic interaction and $|m|$ is the magnetic moment.
Since $\delta J(R)/\delta R\sim0.04$eV/Angstrom
according to the first principle study \cite{Yildirim},
we expect that $\eta_S^{\rm spin}$ is one order of magnitude smaller than 
$\eta_S^{\rm charge}$.
Then, there seems to be difficulty in reproducing 
the softening of $C_S$ that occurs for a wide temperature range
(from the room temperature to $T_S$; see Fig. \ref{fig:CS})
in terms of the spin-nematic scenario.

Finally, we note that the spin-quadrupole scenario
requires that $\chi_{s}(\q)$ is ``commensurate''.
However, recent neutron experiment revealed that
the magnetic order below $T_N$ ($<T_S$) in Ba(Fe$_{1-x}$Co$_x$)$_2$As$_2$
is incommensurate for $x\ge5.7$\%,
although structure transition occurs for $x\ge 7$\%.
This fact would support the realization of two-orbiton mechanism,
since it does work even if $\chi_{xz(yz)}^Q(\q)$ is incommensurate.

\section{Impurity-induced nematic order}
\label{sec:nematic}

\subsection{mean-field approximation}

In previous sections, we have shown that 
both $s_{++}$ wave superconductivity as well as the structure transition
originate from the AF orbital fluctuations,
$\chi_{xz(yz)}^Q({\bm Q})$ caused by the inter-orbital nesting.
In this section, we discuss the nematic ordered state
in the tetragonal phase, which had been observed in many pnictides.
For example, in Ba(Fe$_{1-x}$Co$_x$)$_2$As$_2$, 
large in-plane anisotropy starts to occur
about 10$\sim$100K higher than $T_S$ \cite{detwin}.
This ``nematic electronic state'' free from 
lattice deformation had been also observed by ARPES measurements
\cite{Shimo-Science,ARPES2,ARPES3}, 
optical conductivity measurements \cite{optical},
and the anisotropy in the magnetic susceptibility \cite{Matsuda}.
The origin of this nematic order state
had been one of the great open issues in iron-based superconductors.

Recently, local-density-of-states (LDOS) around impurity sites
had been studied in detail by STM/STS technique,
and nontrivial breakdown of the four-fold symmetry ($C_{4v}$) 
had been found in many pnictides.
For example, single Co impurity induces a remarkable non-local change 
in the DOS along the $a$-axis (longer axis) with a length of 
$\sim 8a_{\rm Fe-Fe}$ in under-doped Ba(Fe$_{1-x}$Co$_x$)$_2$As$_2$
 \cite{Davis}.
Such a drastic impurity-induced anisotropic state
is nontrivial since the orbital polarization in the orthorhombic
phase is very small ($n_{xz}-n_{yz}\sim O(10^{-2})$).
Similar impurity-induced non-local change in the DOS was also observed 
in LiFeAs with tetragonal structure by Hanaguri {\it et al} \cite{Hana}.
In many cases, impurity-induced states break the $C_{4v}$-symmetry,
belonging to $C_{\rm 2v}$, $C_{2}$, or $C_{1h}$.

It would be natural to expect that impurity-induced low symmetric state 
explains the nematic phase above $T_S$.
Recently, in-plane resistivity had been measured in detwinned 
Ba(Fe$_{1-x}$Co$_x$)$_2$As$_2$ crystals with high quality 
\cite{Eisaki,Uchida}.
They found that $\rho_b-\rho_a$ is proportional to 
the Co concentration $x$ for $x\le 4$\%.
Moreover, the residual resistivity per 1\% Co impurity 
reaches $\sim100\mu\Omega$cm, which exceeds the unitary limit
of the local impurity potential ($\sim20\mu\Omega$cm).
Therefore, it had been proved experimentally that a Co impurity 
works as a strong scatter in the FeAs plane,
like Zn impurity in cuprate superconductors.

In strongly correlated electron systems,
impurity potential frequently causes nontrivial non-local change 
in the electronic state.
For example, in nearly antiferromagnetic metals, antiferromagnetic 
correlation is drastically enhanced near the nonmagnetic impurity site,
and the local magnetic moment emerges around the impurity.
Such phenomena are indeed observed in under-doped cuprates
 \cite{Kontani-review}.
As for the iron-based superconductors,
the system would be close to antiferro-orbital QCP.
Therefore, it is natural to expect the occurrence 
of ``impurity-induced non-local orbital order'' in pnictides.

In this section, we study the single impurity problem
with potential $I$ in ten-orbital tight-binding model for pnictides,
using the mean-field approximation in real space
Since we concentrate on the impurity-induced orbital-ordered state,
we take only the quadrupole-quadrupole
interaction in eq. (\ref{eqn:Hquad}) into account:
We consider that the coupling constant $g$ in eq. (\ref{eqn:Hquad})
originates from both the $e$-ph interaction as well as the 
multiorbital Coulomb interaction,
as discussed in Sec. \ref{sec:orbital}.
We obtain the following mean-field self-consistently for $g<g_c$,
where $g_c=0.221$ is the critical value for the bulk orbital order
\cite{Inoue}:
\begin{eqnarray}
M_{l,m}^i= \langle c_{i,l\s}^\dagger c_{i,m\s}\rangle_{I,g}
-\langle c_{i,l\s}^\dagger c_{i,m\s}\rangle_{I,0}
\label{eqn:M}
\end{eqnarray}
where $i$ is the Fe site, 
and $l,m$ represent the $d$-orbital.
Note that $M_{l,m}^i$ is impurity-induced mean-field for $g<g_c$
since it vanishes when $I=0$.
Then, the mean-field potential is given as
\begin{eqnarray}
S_{l,m}^{i}= 2\sum_{l',m'}\Gamma^c_{lm,l'm'}M_{l',m'}^i
\label{eqn:Sigma}
\end{eqnarray}
and the mean-field Hamiltonian is
${\hat H}_{\rm MF}={\hat H}_0+\sum_i{\hat S}^i+ {\rm const}$.
$\Gamma^c$ is the bare four-point vertex for the charge sector.
We solve eqs. (\ref{eqn:M})-(\ref{eqn:Sigma}) self-consistently.

Since the present mean-field has 15 components at each site,
it is convenient to consider the following quadrupole order:
\begin{eqnarray}
O_{\Gamma}^i= 2\sum_{l,m} o_{\Gamma}^{l,m}M_{l,m}^i
\end{eqnarray}
where $\Gamma=xz$, $yz$, $xy$, $z^2$, and $x^2-y^2$.
$o_{\Gamma}^{l,m}$ is the matrix element of the charge quadrupole operator
given in Ref. \cite{Saito}.
We note that the hexadecapole ($l=4$) order is negligible
in the present study.

\begin{figure}[!htb]
\includegraphics[width=.99\linewidth]{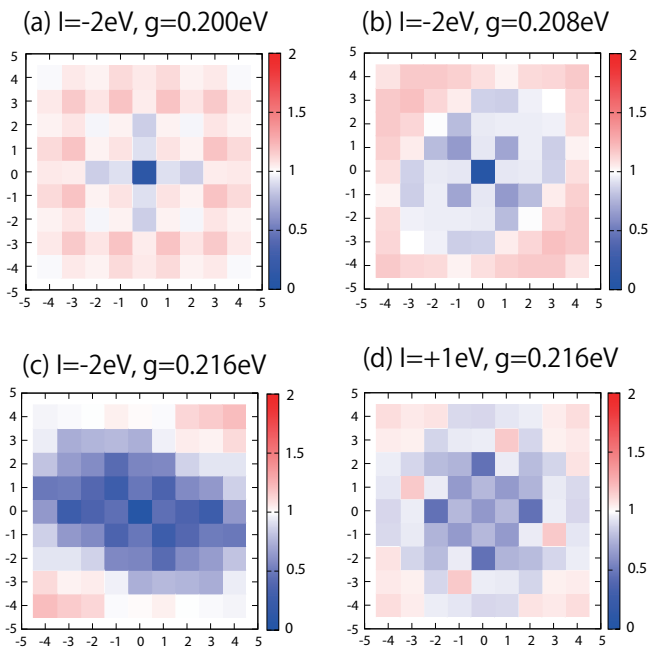}
\caption{
Obtained LDOS given by the mean-field theory
in the presence of impurity potential at $(0,0)$.
(a) $(I,g)=(-2,0.200)$: without orbital order.
(b) $(I,g)=(-2,0.208)$: orbital order with $C_{\rm 2v}$-symmetry.
(c) $(I,g)=(-2,0.216)$: orbital order with $C_{2}$-symmetry.
(d) $(I,g)=(+1,0.216)$: orbital order with $C_{4}$-symmetry.
}
\label{fig:LDOS}
\end{figure}

In Fig. \ref{fig:LDOS}, we show the obtained DOS in real space 
for the cluster of 800 Fe sites with a single impurity site
\cite{Inoue}.
First, we consider the case $I=-2$eV in (a)-(c).
For $g=0.200$ (a), the impurity induced mean-field is absent.
The small modulation of the LDOS around the impurity
is caused by the Friedel oscillation.
For $g\ge0.206$, impurity induced non-local orbital order is induced.
The symmetry of the corresponding LDOS is $C_{\rm 2v}$, 
as shown in (b) for $g=0.210$.
The suppression of the DOS is caused by the non-local orbital order,
consistently with a recent optical conductivity measurement \cite{optical}.
The LDOS are further suppressed with increasing $g$,
and the symmetry is lowered to $C_2$ for $g>0.212$.
In (c), we show the numerical result for $g=0.216$ with $C_2$-symmetry.
Similar impurity-induced LDOS with $C_{2(V)}$-symmetry
is also obtained for a positive impurity potential.
In contrast, we obtain the orbital order with $C_4$ symmetry
for $I=+1$eV.

The dominant impurity-induced quadrupole orders are antiferro
$O_{xz}$, $O_{yz}$ and $O_{xy}$, reflecting the quadrupole-quadrupole
interaction in eq. (\ref{eqn:Hquad}).
The indispensable ingredient for $C_2$-order 
is the quadrupole interaction for $\Gamma=xz/yz$ channels.
We have verified that the quadrupole interaction for 
$\Gamma=x^2-y^2$ channel, which is caused by acoustic phonon, 
cannot realize the $C_2$-order.

By symmetry, the obtained $C_2$-order can be aligned by 
the strain-induced quadrupole potential, which is given in 
eq. (\ref{eqn:HS}) with ${\hat \phi}_S=O_{x^2-y^2}$.
The effective quadrupole potential is 
$\Delta E =\eta_S\e_S\cdot \chi_{x^2-y^2}^Q(0)/\chi_{x^2-y^2}^{(0)}(0)$,
which is strongly enhanced near $T_S$ due to the two-orbiton process
as discussed in Sec. \ref{sec:TO}.
This would be the reason why the nematic ordered state
is easily detwinned by small uniaxial pressure near $T_S$.
In fact, detwinning by uniaxial pressure is possible
only when the structure transition is the second-order \cite{Proz},
in which case the factor $\chi_{x^2-y^2}^Q(0)/\chi_{x^2-y^2}^{(0)}(0)$
develops divergently.
In over-doped systems, 
detwinning by pressure would be difficult 
since $\chi_{x^2-y^2}^Q(0)/\chi_{x^2-y^2}^{(0)}(0)$ is no more large
 \cite{Kontani-soft}.

\subsection{in-plane anisotropy of resistivity}

\begin{figure}[!htb]
\includegraphics[width=.99\linewidth]{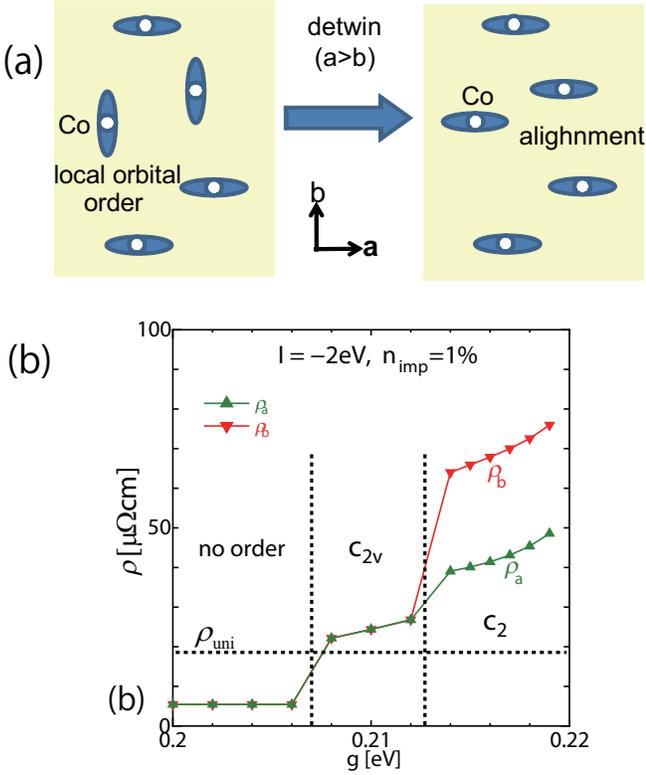}
\caption{
(a) Detwinning of the impurity-induced $C_2$-orders
under uniaxial pressure.
The strain-induced effective quadrupole potential on each Fe is
$\eta_S\e_S (\chi_{x^2-y^2}^Q(0)/\chi_{x^2-y^2}^{(0)}(0)){\hat O}_{x^2-y^2}$.
(b) In-plane anisotropy of resistivity in the nematic phase
obtained by the $T$-matrix approximation.
We assume 1\% impurity with $I=-2$eV.
}
\label{fig:detwin}
\end{figure}

According to ARPES measurements in detwinned systems,
$\Delta E<0$ for $x=a$-axis is longer than $y=b$-axis,
{\it i.e.}, $n_{xz}>n_{yz}$ \cite{ARPES2,ARPES3}.
Then, experimentally observed $(\pi,0)$ SDW order
should be induced theoretically \cite{Kontani-soft}.
In this case, the $C_2$ order aligns along the $x=a$-axis 
in the present numerical results in Figs. \ref{fig:LDOS} (c) and (d).
We have calculated the in-plane resistivity $\rho_a$ and $\rho_b$
in the presence of dilute $C_2$ orders along $a$-axis
using the $T$-matrix approximation, which gives the exact result
when the impurity concentration is dilute and localization is negligible.
We find that $\rho_a$ is smaller than $\rho_b$ by $\sim40$\%,
consistently with experimental reports.
In Fig. \ref{fig:detwin} (a),
we show the alignment of impurity-induced $C_2$ objects
under uniaxial pressure.
The anisotropy of resistivity in the nematic phase
given by the $T$-matrix approximation is shown in Fig. \ref{fig:detwin} (b).

Here, we explain how to calculate the resistivity
in the presence of aligned non-local orbital orders (nematic phase).
The $T$-matrix is given by solving the following equation
in the orbital-diagonal basis:
\begin{eqnarray}
{\hat T}_{{\bm r},{\bm r}'}(\w)&=&({\hat I}+{\hat S})_{\bm r}
\delta_{{\bm r},{\bm r}'}
 \nonumber \\
& &+ \sum_{{\bm r}''}({\hat I}+{\hat S})_{\bm r}
{\hat G}^{(0)}_{{\bm r}-{\bm r}''}(\w){\hat T}_{{\bm r}'',{\bm r}'}^{\rm R}(\w)
 \label{eqn:T}
\end{eqnarray}
where ${\hat S}$ is the mean-field potential,
and ${\hat G}^{(0)}_{\bm r}(\w)$ is the 
Green function in real space without impurities.
${\hat I}_{\bm r}=I{\hat 1}\delta_{{\bm r},{\bm 0}}$ is the impurity potential.
Note that ${\hat S}_{\bm r}$ is diagonal with respect to the 
position since our interaction Hamiltonian is local.

The preset $T$-matrix is non-local because of 
${\hat S}_{\bm r}$ for ${\bm r}\ne{\bm 0}$.
After the Fourier transformation, the self-energy 
due to multiple scattering in the $T$-matrix approximation is
${\hat \Sigma}(\k,\w)= n_{\rm imp}T_{\k,\k}(\w)$,
and the full Green function is
${\hat G}(\k,\w)=(\w+\mu-{\hat H}_\k^0-{\hat \Sigma}^{\rm R}(\k,\w))^{-1}$.
Then, the in-plane conductivity is given by
\begin{eqnarray}
\sigma_\nu=\frac{e^2}{\pi}\sum_{\k,\a}
{v}_{\k,\nu}^\a{J}_{\k,\nu}^\a|G_\a(\k,i\delta)|^2
\end{eqnarray}
where $\nu=x,y$, and $\a$ represents the $\a$th band.
${v}_{\k,\nu}$ is the group velocity of the $\a$th band,
and $G_\a(\k,\w)$ is the full Green function in the band-diagonal basis.
${J}_{\k,\nu}$ is the total current including the current vertex correction,
which is given by solving the following Bethe-Salpeter equation:
\begin{eqnarray}
J_{\k,\nu}^\a = v_{\k,\nu}^\a
 +\sum_{\p,\b} I_{\k,\p}^{\a,\b}|G_\b(\p,i\delta)|^2 J_{\p,\nu}^\b
\end{eqnarray}
where $I_{\k,\k'}=n_{\rm imp}|T^{\rm R}_{\k,\k'}(i\delta)|^2$ is the irreducible vertex.
As shown in Fig. \ref{fig:detwin} (b),
we obtain $\rho_b/\rho_a\sim 2$ in the nematic phase:
The anisotropy is enhanced by including the current vertex correction.

Note that the averaged residual resistivity $(\rho_a+\rho_b)/2$
per 1\% impurity with $I=+1$eV or $-2$eV reaches $\sim50\mu\Omega$cm,
which is comparable to the residual resistivity by 1\% Co
impurities in Ba(Fe$_{1-x}$Co$_x$)$_2$As$_2$; $50\sim100\mu\Omega$cm
\cite{Sato-imp,Uchida}.

Here, we have studied the resistivity
due to elastic scattering caused by spatially extended 
effective impurity potentials.
In addition, strong antiferro fluctuations in pnictides
should present various non-Fermi liquid transport phenomena
in the normal state.
In fact, relations $\rho\propto T$ and $R_{\rm H}\propto T^{-1}$ 
were observed in BaFe$_2$(As,P)$_2$ \cite{Matsuda-rho}.
Although such behaviors are frequently ascribed to the evidence of 
spin fluctuations \cite{Kontani-review},
they are also brought by the development of 
antiferro-orbital fluctuations \cite{Onari-FLEX}.

\subsection{Comparison between theory and experiments}

In a pure two-dimension system,
a single nonlocal orbital order in the mean-field approximation might 
disappear by thermal and quantum fluctuations.
In real systems, however, orbital order would be stabilized 
by the correlation between impurities for $n_{\rm imp}\gtrsim1$\%.
Since the AF fluctuations increase as $T$ decreases in real systems,
we can interpret that $g$ monotonically increases
as $T$ decreases in the present mean-field approximation.
Therefore, we expect that impurity-induced non-local orbital order 
is stabilized below the nematic transition temperature $T^*$, 
at which $g(T)$ would be close to $g_c$.

Although orbital order is absent above $T^*$, 
strong orbital fluctuations should appear near the impurity site,
in analogy to the impurity-induced strong magnetic fluctuations
realized in under-doped cuprates \cite{Kontani-Ohno}.
Therefore, inelastic scattering given by impurity-induced 
strong fluctuations causes large ``residual resistivity'' 
even above the impurity-induced Neel temperature \cite{Kontani-Ohno}.
By the same reason, residual resistivity in under-doped pnictides 
would be large even above $T^*$.
The order parameters for the impurity induced nematic order 
are antiferro-$O_{xz/yz/xy}$, while the order parameter for the 
structure transition is ferro-$O_{x^2-y^2}$.
Because of the difference in order parameters,
$T_S$ would be rather insensitive to the presence of impurities.

The nematic transition also occurs in the tetragonal phase of 
BaFe$_2$(As$_{1-x}$P$_x$)$_2$, which was confirmed 
by the in-plane anisotropy in the magnetization 
($\chi_a\ne\chi_b$) using the torque measurement under magnetic field
\cite{Matsuda}.
In BaFe$_2$(As$_{1-x}$P$_x$)$_2$, $P$ sites would works as impurities, 
which give finite potential on the neighboring for Fe sites.
In this case, we had verified that non-local orbital order with 
$C_{2}$- or $C_{1h}$-symmetry appears around the impurity site 
in the mean-field approximation.
In contrast, nematic order had not be observed in (K,Ba)Fe$_2$As$_2$
\cite{no-nematic},
maybe because impurities outside of FeAs plane would be too weak
to induce orbital order.

In this paper, we consider that the charge quadrupole order
occurs at $T_S$, and the nematic order at $T^* \ (>T_S)$
originates from the impurity-induced orbital order.
However, difference scenario had been proposed by the authors in 
Ref. \cite{Matsuda}:
They consider that the quadrupole order occurs at $T^*$,
and $T_S$ is just a meta-transition without symmetry breaking.
In their phenomenological Landau model,
very small orthorhombicity $(a-b)/(a+b)$ occurs below $T^*$, and it
increases drastically below $T_S$ as a meta-transition, 
consistently with experiments.

\section{Discussion}
\label{sec:summ}

In the present paper,
we have studied the realistic five-orbital model for iron pnictides.
It was found that the $O_{xz}$-AFQ fluctuations develop
owing to the quadrupole-quadrupole interaction 
due to $e$-ph and Coulomb interactions.
This fluctuations not only cause the $s_{++}$-wave superconductivity,
but also the orthorhombic structure transition
due to two-orbiton process.
Using the two-orbiton term,
we can fit the recent experimental data of $C_{S}$ in 
Ba(Fe$_{1-x}$Co$_x$)$_2$As$_2$ \cite{Yoshizawa} for wide range of doping, 
only by choosing $T_{\rm AFQ}$ while other parameters are almost fixed.
This fact is a strong evidence for the success of orbital fluctuation theory
in iron pnictide superconductors.

However, we consider the impurity effect on the $s_{++}$-wave state:
In the weakly correlated metals, impurity effect on 
the $s_{++}$-wave state is very small, known as the 
Anderson theorem.
In strongly correlated metals, in contrast,
impurity-induced change in the many-body electronic states
could violate the Anderson theorem.
In Sec. \ref{sec:nematic},
we have shown that the impurity-induced orbital order
causes remarkable reduction in the DOS,
accompanied by the suppression of orbital fluctuations.
For this reason, orbital-fluctuation-induced $s_{++}$ wave state 
should be suppressed by impurities in the nematic state.
In Ba(Fe$_{1-x}$Co$_x$)$_2$As$_2$,
the suppression in $T_{\rm c}$ per 1\% Zn-impurity,
which gives a strong impurity potential, is $-\Delta T_{\rm c}/\% \sim3$K
 \cite{Li},
while $-\Delta T_{\rm c}/\% \sim 20$K is expected in the $s_\pm$-wave state
when the mass-enhancement is $m^*/m_b \sim 3$
 \cite{Onari-impurity}.
Such small impurity-induced suppression of $T_{\rm c}$ in Ref. \cite{Li}
would be naturally explained by the impurity-induced orbital order.

\begin{figure}[!htb]
\includegraphics[width=.99\linewidth]{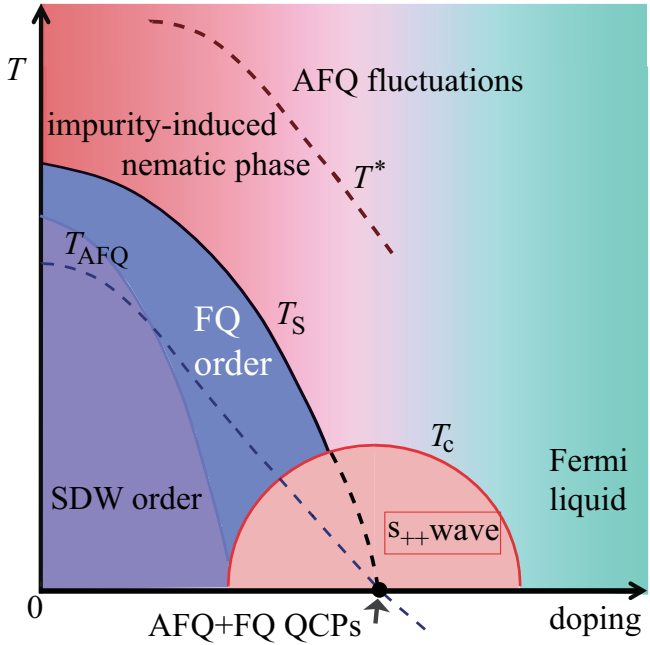}
\caption{
The phase-diagram for iron-pnictide superconductors
obtained by the present orbital fluctuation theory.
$T_S$ is the orthorhombic transition temperature 
(= FQ order temperature), and $T_{\rm N}$ is SDW transition temperature.
The fact that two QCPs at $T_S=0$ and $T_{\rm AFQ}=0$ almost coincide
means that novel ``multi orbital QCPs'' are realized in iron pnictides.
At $T_{\rm AFQ}$, the AFQ-order does not occur 
since it is prevented by the FQ order at $T_S$.
The impurity-induced nematic state is realized below $T^*$.
}
\label{fig:summary}
\end{figure}

In Fig. \ref{fig:summary}, we summarize the phase-diagram
of iron-pnictides given by the present orbital fluctuation theory.
We stress that $T_{\rm AFQ}$, which is determined experimentally from $C_{S}$,
is positive in the under-doped case ($T_S>0$) while it is negative
in the over-doped case.
Especially, $T_{S}\approx0$ for $T_{\rm AFQ}=0$,
consistently with experiments \cite{Yoshizawa,Goto}.
This result indicates that QCPs for AFQ and FQ orders almost coincide
at the endpoint of the orthorhombic phase.
Therefore, AFQ-QCP is not hidden inside of the orthorhombic phase,
favorably to the orbital-fluctuation-mediated $s_{++}$-wave SC state
 \cite{Kontani-RPA,Saito,Onari-FLEX}.
Below $T^*$, impurity-induced orbital order 
with $C_{2(V)}$-symmetry starts to occur.
Each orbital order can be aligned by applying tiny uniaxial pressure,
resulting in the large in-plane anisotropy in the resistivity.

In summary, the present orbital fluctuation theory
can explain the (i) superconductivity, (ii) orthorhombic transition
accompanied by large softening of $C_{S}$, 
and (iii) impurity-induced nematic order.
To explain (i)-(iii), we have to assume that 
$\chi_{xz(yz)}^Q$ is the most divergent quadrupole susceptibility,
which is actually satified in the present model Hamiltonian.
We have derived the orbital fluctuation parameters from $C_{S}$,
and succeeded in explaining $T_{\rm c}$ and $\rho(T)$
using the derived parameters.
These results are strong evidence for the realization of the 
orbital-fluctuation-mediated $s_{++}$-wave superconductivity 
in iron pnictides.

\vspace{5mm}
{\bf acknowledgements}

We are grateful to D.S. Hirashima, 
D.J. Scalapino, P. Hirschfeld and A.V. Chubukov for 
valuable discussions on theories.
We are also grateful to M. Sato, Y. Kobayashi, 
Y. Matsuda, T. Shibauchi, S. Shin, T. Shimojima and M. Yoshizawa
for useful discussions on experiments.
This study has been supported by Grants-in-Aid for Scientific 
Research from MEXT of Japan, and by JST, TRIP.
Numerical calculations were performed using the facilities of 
the supercomputer center, Institute for Molecular Science.





\bibliographystyle{model1-num-names}
\bibliography{<your-bib-database>}



\end{document}